\documentstyle[preprint,revtex]{aps}
\begin{document}
\draft
\input{psfig}
\def\rslide#1#2{\centerline{\hbox{\psfig{figure=#1,height=#2,bbllx=-774bp,bblly
=28bp,bburx=-18bp,bbury=514bp,clip=}} }}
\def\raxum#1#2{\centerline{\hbox{\psfig{figure=#1,height=#2,bbllx=72bp,bblly=-5
30bp,bburx=720bp,bbury=-134bp,clip=}} }}
\def\rtext#1#2{{\hbox{\psfig{figure=#1,height=#2,bbllx=95bp,bblly=769bp,bburx=2
50bp,bbury=783bp,clip=}} }}
\preprint{RUB-TPII-22/92 \\}
\preprint{November, 1992}
\begin{title}
ANALYSIS OF $\Delta^{+}(1232)$ ISOBAR OBSERVABLES \\
WITH IMPROVED QUARK DISTRIBUTION AMPLITUDES
\end{title}
\author{N. G. Stefanis and M. Bergmann}
\begin{instit}
Institut f\"ur Theoretische Physik II  \\
Ruhr-Universit\"at Bochum  \\
D-4630 Bochum, Germany
\end{instit}

\begin{abstract}
A model distribution amplitude for the  $\Delta^{+}(1232)$ isobar
is proposed, derived on the basis of the QCD sum-rule calculations
of Farrar et al. combined with those of Carlson and Poor. The
transition form factor
$\gamma p\Delta^{+}$ is calculated modeling the proton by selected
distribution amplitudes. Furthermore, predictions are made for
some exclusive charmonium decays into $\Delta\bar{\Delta}$.
The obtained results are compared with other theoretical models
and with the available data.
\end{abstract}
\pacs{Internet: michaelb@photon.tp2.ruhr-uni-bochum.de \\
\phantom{Internet:\ }nicos@hadron.tp2.ruhr-uni-bochum.de \\
Bitnet: KPH509@DJUKFA11  }
\narrowtext
\newpage  
The short-distance behavior of exclusive amplitudes and in particular
of the hadronic form factors at large momentum transfer has attracted
continuous interest in the past decade. The theoretical basis is
provided by the factorization theorem within a convolution formalism
\cite{LB80}. The nonperturbative information on the quark structure
of the hadron is contained in the distribution amplitude
$\Phi(x_{i},Q^{2})$, which is the integral over the transverse
momenta of the hadron valence-quark wave function. In a physical gauge
(e.g. $A^{+}=0$), it is the probability amplitude for the hadron to
consist of valence quarks with longitudinal momentum fractions
$0\le x_{i}\le 1$, $\sum_{i}x_{i}=1$ (in a $p_{\infty}$ frame)
moving collinearly up to a scale $Q^{2}$. Contributions from higher Fock
states are suppressed by powers of the momentum transfer.

Theoretical constraints on the shape of the hadron distribution
amplitudes can be obtained from QCD sum rules~\cite{CZ84a} and lattice
gauge theory~\cite{RSS87}. For the nucleon, various model distribution
amplitudes have been proposed~\cite{CZ84b,GS86,KS87,COZ89a} and
detailed studies have been carried out~\cite{JSL87,CP87a,St89,COZ89b}.
More recently, the quark structure of the $\Delta^{+}(1232)$
and higher resonances has been investigated~\cite{Ca86,WC90,Sto91}
and models for the valence part of their distribution amplitudes
have been worked out~\cite{CP88,FZOZ88}.

It was shown in~\cite{CGS87} that
the transition form factor $G_{M}^{*}$ and the neutron
magnetic form factor $G_{M}^{n}$ are anti-correlated so that one
of those form factors tends to be large when the other is small.
If one uses the CZ~\cite{CZ84b} or the COZ~\cite{COZ89a} nucleon
distribution amplitude, $\vert G_{M}^{*}\vert $ turns out to be nearly
zero, whereas ${\vert G_{M}^{n}\vert}/{G_{M}^{p}}\le 0.5$.
This possibility would explain why the data on the ratio
${F_{E2}}/{F_{M1}}$ yield a value compatible with zero.
On the other hand, the GS~\cite{GS86} nucleon distribution amplitude
gives by construction ${\vert G_{M}^{n}\vert}/{G_{M}^{p}}\le 0.1$
and a $\vert G_{M}^{*}\vert $ of the same order of magnitude as
$G_{M}^{p}$. This behavior is consistent with the possibility that
$\sigma_{n}$, the electron-neutron differential cross section, is
dominated by $G_{E}^{n}$, while $G_{M}^{n}$ is asymptotically small or
equivalently that $\vert F_{1}^{n}\vert\ll\vert F_{2}^{n}\vert$ at all
$Q^{2}$ values~\cite{KK77}. These results were obtained under the
assumption that the $\Delta$ distribution amplitude can be approximately
represented by the symmetric part of the nucleon one. However, the
anti-correlation pattern still holds also for more realistic models for
the $\Delta$ distribution amplitude (see~\cite{CP88} and below).

{}From the experimental side~\cite{DATD}, there is still no clear evidence
whether $Q^{4}\vert G_{M}^{*}\vert$ levels off or drops to zero at large
$Q^{2}$. A recent analysis~\cite{Sto91} of the unpublished data of the
SLAC experiment E133 points rather to the second possibility, though the
precision of the data is yet not sufficient to single out one
possibility over the other.

The theoretical predictions for the normalization and the $Q^{2}$
evolution of $G_{M}^{*}$ are sensitive to the shape of the proton
distribution amplitude.
It is therefore essential to have reliable models for both the
nucleon and the $\Delta$ distribution amplitudes to be able to make
comparisons with the data in precise detail.

In a previous letter~\cite{SB92}, we have shown that two widely used
models for the nucleon distribution amplitude,
the Chernyak-Ogloblin-Zhitnitsky (COZ) model and
the Gari-Stefanis (GS) model --- treated previously
as competing alternatives --- can be unified into a single model: the
heterotic model. This model provides the possibility to analyze the data
on $\sigma_{n}$ under the assumption that
${\vert G_{M}^{n}\vert}/{G_{M}^{p}}\le 0.1$
(like the GS model), while matching the sum-rule requirements
\cite{COZ89a} on the moments of $\Phi_{N}$ up to the third order with
almost the same overall accuracy as with the COZ model. Furthermore,
the calculated decay widths of the charmonium states
$^3S_{1}$, $^3P_{1}$, and $^3P_{2}$ into $p \bar p$ are in excellent
agreement with the data~\cite{PDG90} without individual adjustment of the
various parameters. No other existing model is so successful in that
respect~\cite{COZ89b,BS92,ACF91}.

In the present letter, we apply similar ideas to derive an optimum
distribution amplitude for the $\Delta^{+}(1232)$ isobar.
The new element of our approach is that we treat the sum-rule analysis
of Carlson and Poor~\cite{CP88} in conjunction with that of
Farrar et {\it al.}~\cite{FZOZ88}. A second result of this paper
is the calculation of the transition form factor $G_{M}^{*}$ in
remarkable agreement with the data. In addition, we make predictions for
the decay widths of the charmonium states
$^3S_{1}$, $^3P_{1}$, and $^3P_{2}$ into $\Delta \bar \Delta$.

To leading order, the quark distribution amplitude for the baryon with
helicity $+1/2$ can be represented in the form:
\mediumtext
\begin{equation}
  \Phi_{[1/2]}(x_{i},Q^{2}) =
\Phi_{as}(x_{i})\sum_{n=0}^{\infty}B_{n}^{[1/2]}(\mu^{2})
  \Biggl({{\alpha_{s}(Q^{2})}\over{\alpha_{s}(\mu^{2})}}
\Biggr)^{\gamma_{n}}
  \tilde \Phi_{n}(x_{i})\;,   \label{evoleq}
\end{equation}
\narrowtext
where \hbox{$\Phi_{as}(x_{i})=120\, x_{1}x_{2}x_{3}$} and
$\{\tilde \Phi_{n}(x_{i})\}$ are the eigenfunctions of the
interaction kernel of the evolution equation, expressed
in terms of Appell polynomials~\cite{LB80}.
The corresponding eigenvalues $\gamma_{n}$ equal the anomalous
dimensions of the lowest-twist three-quark operators carrying the
appropriate baryonic quantum numbers and are perturbatively calculable
renormalization-group coefficients~\cite{Pe79}. Note that in the
$\Delta$ case only eigenfunctions symmetric under
\hbox{$x_{1}\leftrightarrow x_{3}$}
contribute (i.e. $B_{1}^{\Delta} = B_{4}^{\Delta} \equiv 0$).

The expansion coefficients $B_{n}^{[1/2]}$ are nonperturbative quantities
representing matrix elements of three-quark operators interpolating
between the vacuum and the helicity $+1/2$ baryon (the nucleon (N) or
the $\Delta^{+}(1232)$ isobar), renormalized at a scale~$\mu^{2}$:
\mediumtext
\begin{equation}
  <\Omega\vert O_{\gamma}^{(n_{1}n_{2}n_{3})}(0)\vert b^{[1/2]}(p)>
  =
  f_{N(\Delta)}(z\cdot p)^{\,{n_{1}+n_{2}+n_{3}+1}}S_{\gamma}^{[1/2]}
(p)  \,  O^{\,(n_{1}n_{2}n_{3})}.
\end{equation}
\narrowtext
Here $z$ is a lightlike vector ($z^{2}=0$), with $z\cdot q=
q^{+}=q^{0}+q^{3}$ for any vector $q$, $S_{\gamma}^{[1/2]}(p)$
is the spin function of the baryon with helicity $+1/2$, and
$f_{N(\Delta)}$ is a dimensionful constant denoting the value of the
matrix element at the origin.

To carry out sum-rule
calculations~\cite{CZ84a,CZ84b,KS87,St89,CP88,FZOZ88},
one uses correlators involving two of the
$O_{\,\gamma}^{(n_{1}n_{2}n_{3})}$:
\mediumtext
\begin{eqnarray}
I^{\,(n_{1}n_{2}n_{3},m)}(q,z) & = & i\int_{}^{}
  d^{4}x \, e^{iq\cdot x}
  <\Omega\vert T\bigl (O_{\gamma}^{\,(n_{1}n_{2}n_{3})}(0)
  \hat O_{\gamma\prime}^{\,(m)}(x)\bigr )\vert\Omega>(z\cdot \gamma)_
  {\gamma \gamma\prime} \nonumber\\
   &  = &
  (z\cdot
  q)^{\,{n_{1}+n_{2}+n_{3}+m+3}}I^{\,(n_{1}n_{2}n_{3},m)}(q^{2})\;,
\end{eqnarray}
\narrowtext
where $(z\cdot\gamma)_{\gamma\gamma\prime}$ serves to project out the
leading twist-structure of the correlator. To obtain constraints on the
moments of the baryon distribution amplitude,
\begin{equation}
  \Phi_{N(\Delta)}^{\,(n_{1}n_{2}n_{3})} = \int_{0}^{1}[dx]\,
  x_{1}^{n_{1}}\,x_{2}^{n_{2}}\,x_{3}^{n_{3}}\,\Phi_{N(\Delta)}(x_{i})\;,
\end{equation}
(\hbox{$[dx]=dx_{1}dx_{2}dx_{3}\,\delta(1-x_{1}-x_{2}-x_{3})$})
a short-distance operator product expansion is performed at some
spacelike momentum $\mu^{2}$ where quark-hadron duality is valid.
By virtue of the orthogonality of the eigenfunctions $\tilde\Phi_{n}$,
the coefficients $B_{n}^{[1/2]}(\mu^{2})$ and the constant
$f_{N(\Delta)}$ can be determined by means of moments inversion using as
constraints the sum-rule requirements. The baryon distribution amplitude
is then expressed in the form of a truncated series of Appell
polynomials of up to second order, which means that we take into account
the first six terms: $n=0,1,\ldots 5$.

The coefficients for the heterotic nucleon distribution amplitude
were determined in~\cite{SB92}:
$B_{0}=1$ (due to the normalization condition
$\int_{0}^{1}[dx]\,\Phi_{N}(x_{i})=1$
), $B_{1}=3.4437$, $B_{2}=1.5710$,
$B_{3}=4.5937$, $B_{4}=29.3125$, and $B_{5}=-0.1250$.
[Here and below the
conventions and analytical expressions given in~\cite{St89} are used.]
The profile of $\Phi_{N}^{Heterotic}$ is shown in Fig.~\ref{Figr1}.

In this note we treat the sum rules defined in Eq. (3) for
\hbox{$n_{1}+n_{2}+n_{3}\le 3$} and $m=1$ by exploiting the possibility
of simultaneously satisfying the moment constraints on the $\Delta$
distribution amplitude of both the Carlson and Poor~\cite{CP88} and the
Farrar {\it et al.}~\cite{FZOZ88} analyses. In this way we obtain a
$\Delta$ distribution amplitude which has the explicit form
\mediumtext
\begin{eqnarray}
 \Phi_{\Delta}^{Het}(x_{i}) &= \Phi_{as}(x_{i}) \bigl
 \{&  7.2041-8.2859x_{2}-21.3682x_{1}x_{3} 
-4.9247(x_{1}^{2}+x_{3}^{2})\bigr \}\; 
\end{eqnarray}
\narrowtext
with coefficients $B_{n}^{\Delta}$ given in Table~\ref{Tab1} in
comparison with those we determined for the specific model forms
proposed by the above authors. Remarkably, like the nucleon case, this
solution has heterotic character (see Fig.~\ref{Figr2}).

{}From Table~\ref{Tab2} we can see that all the moments of the heterotic
solution are within the range calculated by FZOZ for the amplitude
$T_{\Delta}(x_{1},x_{2},x_{3})=\Phi_{\Delta}(x_{2},x_{3},x_{1})$.
Furthermore, the CP constraints on the amplitude
$V_{\Delta}(x_{1},x_{2},x_{3})=\Phi_{\Delta}(x_{1},x_{2},x_{3})+
A_{\Delta}(x_{1},x_{2},x_{3})$ are also satisfied, with the exception of
the moment $V^{(001)}$ for which the heterotic solution yields a value
slightly smaller than the estimated minimum value of the corresponding
sum rule. Note that if one takes the estimated margins of both
analyses \cite{CP88,FZOZ88} as they stand, then it is not
possible to find a solution for the $\Delta$ distribution amplitude
of the form given by Eq.~\ref{evoleq}
which simultaneously satisfies all sum-rule constraints.

Following~\cite{Ca86} we input the heterotic nucleon distribution
amplitude to calculate the $N-\Delta^{+}$ transition form factor
$G_{M}^{*}$, modeling the $\Delta^{+}$ isobar by the three options
labeled CP, FZOZ, and Heterotic. The $Q^{2}$ evolution is due to the
one-loop approximation of $\alpha_{s}(Q^{2})$ with $\Lambda_{QCD}=180$
MeV~\cite{GS87}. Here the average of two coupling constants
$\bar \alpha_{s}^{N}(Q^{2})$ and $\bar \alpha_{s}^{\Delta}(Q^{2})$
is used, with arguments weighted by the virtualities determined
for each model. For the heterotic $\Delta$ distribution amplitude
we obtain
$\bar \alpha_{s}^{\Delta (Het)}(Q^{2})=[\alpha_{s}(Q^{2}\times 0.494)\,
 \alpha_{s}(Q^{2}\times 0.088)]^{1/2}$; the corresponding values of
$\bar \alpha_{s}(Q^{2})$  for the other considered models are given
in \cite{SB92} and references cited therein.

The results are shown in Fig.~\ref{Figr3}, where also comparison is made
with the predictions for $G_{M}^{*}$ derived from
$\Phi_{\Delta}^{Heterotic}$ in conjunction with previous models for
nucleon distribution amplitudes. The experimental
data are compiled in~\cite{DATD}. In all cases the CP value
$\vert f_{\Delta}\vert = 11.5\times 10^{-3}\ GeV^{2}$
has been used, which is within the spread of the FZOZ estimate.
We emphasize that the sign of $G_{M}^{*}$ predicted by CZ~\cite{CZ84b},
COZ~\cite{COZ89a}, and GS~\cite{GS86} comes out negative for all
$\Delta$ distribution amplitudes discussed here (cf.~\cite{BS92}).
It is only the heterotic~\cite{SB92} nucleon distribution amplitude
and the KS~\cite{KS87} one that yield a positive sign for $G_{M}^{*}$.

In order to account for (unknown) confinement effects at low $Q^{2}$,
we saturate $\alpha_{s}$ by introducing an effective gluon mass:
$\bar \alpha_{s}(Q^{2}) \mapsto
\bar \alpha_{s}(Q^{2} + 4\, m_{g}(Q^{2}))$. The
situation is illustrated in Fig.~\ref{Figr4}. In contrast to other
approaches of this type~\cite{JSL87,HEG92}, we use a dynamical, i.e.,
scale-dependent gluon mass derived by Cornwall~\cite{Co82}:
\mediumtext
\begin{equation}
m_{g}(Q^{2})=m_{g}^{2}\Biggl\{ \ln\Biggl(\frac{Q^{2}+4\, m_{g}^{2}}
{\Lambda_{QCD}^{2}}\Biggr) {\Bigg /} \ln\Biggl( \frac{4\, m_{g}^{2}}
{\Lambda_{QCD}^{2}}\Biggr)\Biggr\}^{-12/11}.
\end{equation}
\narrowtext
Due to the positivity of the anomalous dimension of the mass operator,
this gluon mass vanishes asymptotically. This soft behavior at short
distances leaves the validity of the form-factor evolution at large
momentum transfer virtually unaffected. In the fit shown in
Fig.~\ref{Figr4} (dashed-dotted line), we take $m_{g}=380$ MeV, which
agrees with Cornwall's consistency relation $m_{g}/\Lambda_{QCD}
\approx 1.5 - 2.0$.

Referring to the same figure, we see that including the perturbative
$Q^{2}$ evolution of the coefficients $B_{n}^{\Delta}$ (cf. Eq. (1)),
it is sufficient to provide a good fit to the data above $Q^{2}\approx 3
\ GeV^{2}/c^{2}$ (dashed line). At lower $Q^{2}$ values, additional
nonperturbative parameters have to be introduced in the way just
described (e.g., effective parton masses, quark clustering etc.) to
account for the limitations of the leading-order formalism.

In conclusion of this work, we make predictions for the exclusive decays
of the charmonium levels $^3S_{1}$, $^3P_{1}$, and $^3P_{2}$
into $\Delta \bar \Delta$. We here follow~\cite{COZ89b,SB92}.
The branching ratio of the decay of the
$J^{CP}=1^{++}$ state into $\Delta \bar \Delta$ is given by
\begin{equation}
 BR\Biggl({{^3P_{1}\to \Delta\bar\Delta}\over {^3P_{1}\to all}}\Biggr)
 \approx {{0.75}\over{\ln({\bar M}/{\Delta})}}
 {{16\pi^{2}}\over{729}}{\Bigg \vert {{f_{\Delta}}\over{{\sqrt 3\bar
 M}^{2}}}
 \Bigg \vert}^{4}{(M_{1}^{\Delta})^{2}},
\end{equation}
where $\bar M\approx 2 m_{c} \approx 3$ GeV and $\Delta =0.4$ GeV
(the last value is taken from~\cite{BARB}-see also~\cite{Nov78}).
The analogous expression for the $J^{PC}=2^{++}$ state
has the form
\begin{equation}
 BR\Biggl({{^3P_{2}\to \Delta\bar\Delta}\over {^3P_{2}\to all}}\Biggr)
  \approx 0.85 (\pi\alpha_{s})^{4}
  {{16}\over{729}}{\Bigg \vert {{f_{\Delta}}\over{{\sqrt 3\bar M}^{2}}}
  \Bigg \vert}^{4}{(M_{2}^{\Delta})^{2}}\;.
\end{equation}
The partial width of the $J^{PC}=1^{--}$ state into
$\Delta \bar \Delta$ is
\begin{equation}
  \Gamma(^3S_{1}\to \Delta \bar \Delta) =
   (\pi\alpha_{s})^{6}{{1280}\over{243\pi}}
  {{{\vert f_{\psi}\vert}^{2}}\over{\bar M}}
    {\Bigg \vert {{f_{\Delta}}\over{{\sqrt 3\bar M}^{2}}}
  \Bigg \vert}^{4}{(M_{0}^{\Delta})^{2}}\;,
\end{equation}
where $f_{\psi}$ determines the value of the $^3S_{1}$-state wave
function at the origin. Its value can be extracted from the leptonic
width $\Gamma(^3S_{1}\to e^{+}e^{-}) = (4.72\pm 0.35)$ keV~\cite{PDG90}
via the Van Royen-Weisskopf formula. The result is
$\vert f_{\psi}\vert =383$ MeV with $m_{J/\psi}$ equal
to its experimental value~\cite{PDG90}. The nonperturbative input is due to
$f_{\Delta}$ and the decay amplitudes $M_{1}^{\Delta}$,
$M_{2}^{\Delta}$, and $M_{0}^{\Delta}$. The results for the specific
models considered here are summarized in Table~\ref{Tab3}. For the
branching ratio of the ${}^3S_{1}$ state, the heterotic
distribution amplitude yields
$ BR({{^3S_{1}\to \Delta\bar\Delta} / {^3S_{1}\to all}})
  = 0.30 \times 10^{-2} \% $, where the new \cite{HP92} value
 $\Gamma_{tot} = 85.5^{+6.1}_{-5.8}$ keV is used. Note that
in all considered decays $\alpha_{s}(m_{c}) = 0.210 \pm 0.028$ (see
third paper of \cite{BARB}).

We have performed our analysis pretending that factorization applies to
exclusive reactions, so that a perturbative treatment is justified at
accessible momentum transfer. This issue has been questioned by Isgur
and Llewellyn-Smith~\cite{ILS84} and also by Radyushkin~\cite{Rad91},
who argued that soft contributions dominate even at enormous momentum
scales, rendering perturbative QCD inadequate for exclusive reactions.
However, more recently, Li and Sterman~\cite{LS92} have shown that there
is a an infrared protection of the perturbative picture provided by
Sudakov effect suppression.

In summary, we believe that the heterotic distribution amplitudes
$\Phi_{N}^{Het}$ and $\Phi_{\Delta}^{Het}$ are physically appropriate
solutions in terms of which several exclusive reactions can be
theoretically described without invoking additional assumptions and
without tuning the various physical parameters from case to case.

\figure{Profile of the heterotic nucleon distribution
        amplitude.\label{Figr1}}
\figure{Profiles of the model distribution amplitudes for
        $\Delta^{+}(1232)$ described in the text.\label{Figr2}}
\figure{Comparison with available data of the transition form factor
           $\gamma p \Delta^{+}$
           calculated with the heterotic nucleon distribution
           amplitude and three different model distribution
           amplitudes for the $\Delta^{+}$
           isobar, as explained in the text. The solid horizontal
           lines are predictions for the absolute value of
           $G_{M}^{*}$ at $Q^2\,=\,15\,GeV^{2}/c^{2}$
           of other nucleon distribution amplitudes
           labeled CZ~\cite{CZ84b}, GS~\cite{GS86},
           KS~\cite{KS87}, and COZ~\cite{COZ89a}. The data denoted by
           open circles are from \cite{Sto91}. \hfil
           \label{Figr3}}
\figure{Comparison with available data of the transition form factor
           $\gamma p \Delta^{+}$
           calculated with the heterotic nucleon distribution
           amplitude and the heterotic distribution amplitude
           for the $\Delta^{+}$ isobar. Three different calculations
           are shown:
           the one-loop approximation of $\alpha_{s}(Q^{2})$ (solid
           line), a modified expression for $\alpha_{s}(Q^{2})$
           which takes into account a
           dynamical gluon mass $m_{g}(Q^{2})$ (dashed-dotted line),
           and the effect of $Q^{2}$-evolution of the coefficients
           $B_{n}^{\Delta}$ (dashed line).\hfil
           \label{Figr4}}

\def\za{\phantom{1}}
\def\zb{\phantom{12}}
\def\zf{\phantom{(0.321)}}
\def\zm{\phantom{--}}
\narrowtext
\begin{table}
\caption{Expansion coefficients, $B_{n}^{\Delta}$, of the
         model distribution amplitudes for $\Delta^{+}(1232)$,
         discussed in the text. }
\begin{tabular}{cccc}
         $B_{n}^{\Delta}$ &  CP & FZOZ &  Heterotic \\
\tableline
         $B_{2}^{\Delta}$  & 0.35\zb  & --0.175   & --0.2499  \\
         $B_{3}^{\Delta}$  & 0.4095   &\zm 1.071  &\zm 0.3297 \\
         $B_{5}^{\Delta}$  & 0.1755   & --0.486   & --1.6205  \\
\end{tabular}
\label{Tab1}
\end{table}
\newpage

\begin{table}
\caption{Moments $n_{1}+n_{2}+n_{3}\le 3$ of the model
         distribution amplitudes for $\Delta^{+}(1232)$ in
         comparison with the sum-rule constraints.
         The numbers in parentheses are those given by
         FZOZ~\cite{FZOZ88}.}
\begin{tabular}{ccccc}
         Moments & Sum rules &\multicolumn{3}{c}{Models} \\
         ${(n_{1}n_{2}n_{3})}$ & $T_{\Delta}(FZOZ)$ & CP & FZOZ &
         Heterotic \\
\tableline
         (000)   & 1                & 1     &    1                & 1      \\
         (100)   & 0.31---0.35      & 0.35\za & 0.325 (0.32)\za   & 0.321  \\
         (001)   & 0.35---0.40      & 0.30\za & 0.350 (0.36)\za & 0.357  \\
         (200)   & 0.14---0.16      & 0.16\za & 0.150   \zf       & 0.14\za \\
         (002)   & 0.15---0.18      & 0.123   & 0.160   \zf       & 0.151  \\
         (110)   & 0.07---0.1\za    & 0.101   & 0.080   (0.07)\za & 0.078  \\
         (101)   & 0.09---0.13      & 0.089   & 0.095 (0.1)\zb    & 0.103  \\
         (300)   & 0.06---0.09      & 0.085   & 0.083 (0.085)   & 0.073  \\
         (003)   & 0.06---0.10      & 0.060   & 0.085 (0.081)   & 0.071  \\
         (210)   & 0.025---0.04\za  & 0.039   & 0.030 (0.025)   & 0.027  \\
         (201)   & 0.04---0.06      & 0.035   & 0.037 (0.04)\za & 0.040  \\
         (102)   & 0.035---0.06\za  & 0.031   & 0.037 (0.039)   & 0.040  \\
\tableline
        & $V_{\Delta}(CP)$ &  &  & \\
\tableline
         (001)   & 0.33---0.37      & 0.35\za & 0.325          & 0.321   \\
         (002)   & 0.14---0.18      & 0.16\za & 0.15\za        & 0.14\za \\
         (101)   & 0.072---0.12\za  & 0.095   & 0.088          & 0.091   \\
\end{tabular}
\label{Tab2}
\end{table}
\newpage

\begin{table}
\caption{Charmonium decays in  $\Delta\bar{\Delta}$ for the
         models discussed in the text. }
\begin{tabular}{cccc}
 Amplitudes   &  CP & FZOZ &  Heterotic \\
\tableline
         $M_{1}^{\Delta}$  & 11418.23   &   21585.99  &   11651.26  \\
         $M_{2}^{\Delta}$  & 30924.07   &   48233.23  &   26277.40 \\
         $M_{0}^{\Delta}$  &\za 1480.67 &\za 1882.31  &\za 1134.94 \\
\tableline
 Observables & & & \\
\tableline
$BR\Bigl({{^3P_{1}\to \Delta\bar{\Delta}}\over {^3P_{1}\to all}}\Bigr)$
& $0.311\times 10^{-3}\%\ \ \ $ & $1.113\times 10^{-3}\%\ \ \ $ & $0.324\times
10^{-3}\%\ \ \ $
\vspace{.1cm} \\
$BR\Bigl({{^3P_{2}\to \Delta\bar{\Delta}}\over {^3P_{2}\to all}}\Bigr)$
& $0.100\times 10^{-3}\%\ \ \ $ & $0.106\times 10^{-3}\%\ \ \ $ & $0.072\times
10^{-3}\%\ \ \ $
\vspace{.1cm} \\
$ \Gamma(^3S_{1}\to \Delta\bar{\Delta}) $
& $0.439\times 10^{-2}\, keV$& $0.709\times 10^{-2}\, keV$& $0.258\times 10^{-2
}\, keV$
\\
\end{tabular}
\label{Tab3}
\end{table}
\newpage

\end{document}